# Bug Classification: Feature Extraction and Comparison of Event Model using Naïve Bayes Approach

Sunil Joy Dommati, Ruchi Agrawal, Prof. Ram Mohana Reddy.G and Sowmya Kamath

*Abstract*—In software industries, individuals at different levels from customer to an engineer apply diverse mechanisms to detect to which class a particular bug should be allocated. Sometimes while a simple search in Internet might help, in many other cases a lot of effort is spent in analyzing the bug report to classify the bug. So there is a great need of a structured mining algorithm - where given a crash log, the existing bug database could be mined to find out the class to which the bug should be allocated. This would involve Mining patterns and applying different classification algorithms. This paper focuses on the feature extraction, noise reduction in data and classification of network bugs using probabilistic Naïve Bayes approach. Different event models like Bernoulli and Multinomial are applied on the extracted features. When new unseen bugs are given as input to the algorithms, the performance comparison of different algorithms is done on the basis of accuracy and recall parameters.

## I. INTRODUCTION

WITH the advancement in software technology, as number of software products are increasing, maintenance is becoming a challenging task. Maintenance activities account for over two third of the life cycle cost of a software system [4]. Hence a lot of time and efforts are required for maintenance phase of software development lifecycle. Essential activities involved in maintenance are bug reporting and fixing [1]. Many developers put significant amount of effort for finding and debugging software bugs. In addition, a significant amount of submitted bug reports are duplicates that describe already-reported bugs [2]. Sometimes while a simple search in Internet might help, in many cases a lot of effort is spent in analyzing the bug reports to classify the bug. In practice, considerable number of the duplicate bugs is reported daily; manually labeling these bugs is highly time consuming task.

To address above mentioned issue, several techniques have been proposed using various similarity based metrics. The similarity measures are used to detect candidate duplicate bug reports or identical bugs and to classify bugs [3]. The same bug can be reported in two different ways and hence extracting features and classifying the bug reports become very complicated. Different discriminative model based approaches are also used for classifying [2].

But all these approaches are generalized, which can be applied to any type of bug database and gives an accuracy of around 30-35% [2]. The accuracy can still be improved if bug semantics are taken into consideration. The bug classification is highly dependent on the type and characteristic of the bugs. For example, network bugs have different semantics as compared to bugs related to IDE's (ECLIPSE, Netbeans etc) or bugs related to browsers (Mozilla, Chrome). Feature Extraction based on severity, priority and other general information is independent of the types of bugs. We think that if the feature extraction is done on the basis of the bug specific characteristic then efforts of the developer for bug fixing can be minimized.

In this paper, we analyzed the network bugs and depending on the static analysis of the bug reports, the feature extraction and selection has been performed. Feature Extraction from bug report is performed according to networking protocols, operating system related defects, product related bugs and different networking protocols to which they belong such as Border Gateway Protocol (BGP), Internet Protocol (IPv4 and IPv6), and Transmission Control Protocol (TCP) etc. Different bug specific features are ranked according to Information Gain criteria. Two different event models: Bernoulli and Multinomial Model can be considered for classification.

## II. RELATED WORK

Davor Cubranic and Gail C.Murphy have proposed an approach for automatic bug triage using text categorization [5]. They proposed a prototype for bug assignment to developer using supervised Bayesian learning. The evaluation shows that their prototype can correctly predict 30% of the report assignments to developers. The prototype used the

Sunil Joy.Dommati is with the Information Technology Department, National Institute of Technology, Karnataka, 575025 INDIA (e-mail: suniljoy777@gmail.com).
Ruchi. Agrawal, is with Information Technology Department, National Institute of Technology, Karnataka, 575025 INDIA (e-mail: agrawalruchi01@gmail.com).
Prof. Ram Mohana Reddy .G, is with the Information Technology Department, National Institute of Technology, Karnataka,575025 INDIA, (e-mail: profgrmreddy@gmail.com ).
Mrs. Sowmya.Kamath, is with the Information Technology Department, National Institute of Technology, Karnataka, 575025, INDIA,(e-mail: sowmyakamath@gmail.com ).

word frequency as input to the classifier. The words were extracted using Natural Language Processing Techniques. The words can be considered as unigram features obtained irrespective of the type of bugs. They analyzed their technique on open source eclipse bug database.

Nicholas Jalbert and Westley Weimer have proposed a system that automatically classifies duplicate bug reports as they arrive to save developer time [6]. Their system used surface features, textual semantics, and graph clustering to predict duplicate status. Using a dataset of 29,000 bug reports from the Mozilla project, they performed experiments that include a simulation of a real-time bug reporting environment. Their system was able to reduce development cost by filtering out 8% of duplicate bug reports while allowing at least one report for each real defect to reach developers.

Deqing Wang, Mengxiang Lin, Hui Zhang, Hongping Hu have implemented a tool Rebug-Detector, to detect related bugs using bug information and code features[7]. They extracted features related to bugs and used relationship between different methods that is overloaded or overridden methods. They evaluated Rebug-Detector on an open source project: Apache Lucene-Java. The results show that bug features and code features extracted by their tool are useful to find real bugs in existing projects.

The classification of the bugs into different buckets can be done using data mining and machine learning concepts. The bugs will be classified into different buckets according to selected features. The approach of bucketing was used by Microsoft for their Windows Error Reporting System [8]. Windows Error Reporting (WER) is a distributed system that automates the processing of error reports coming from an installed base of a billion machines. It collects error data automatically and classifies errors into buckets, which are used to prioritize developer effort and report fixes to users. For Bucketing two types of heuristics were applied: Expanding heuristics increase the number of buckets with the goal that no two bugs are assigned to the same bucket, Condensing heuristics decrease the number of buckets with the goal that no two buckets contain error reports from the same bug. The two heuristics are complementary to each other and ensure the correct and efficient bucketing.

Karl-Michael Schneider in the paper[10] used Naive Bayes Method for Spam Classification. Kian Ming Adam Chai, Hwee Tou Ng and Hai Leong Chieus in their paper [9], explores the use of Bayesian probability approach for text classification. They had used ltc normalization and compared two different types of Bayesian classifiers that is Bayesian Online perception and Bayesian Gaussian Process. They showed through experiments that Bayesian is good approach for text classification.

III. FEATURE EXTRACTION AND SELECTION

A. Overiew of the Bug Site

In bug site, bug reports are organized in the form of different attachments and attachments are grouped into General, Commit, Build, Test, Fix Entries category. According to us, attachments of General category are relevant for classification purpose. General category attachments contain information which is available before the bug is analyzed, tested and fixed by the developer. General category attachments are further divided into Description, Crash info, Decode file, Event log, Email, Static analysis etc attachments. Then bug information is extracted by analyzing the attachments and irrelevant attachments are discarded. For example, Static analysis and Email information etc. are discarded from General category. The information is retrieved from the bug site in html format; html tags are then removed to get individual paragraphs. Information is then statically analyzed to find some pattern for automatic feature extraction.

B. Feature Extraction and Preprocessing

The goal of a bug feature extractor is to automatically extract features from bug information in bug repository after html tag removal. That is, to extract bug information from attachments written in natural language and from the programming language information present in Crash info attachment. Developers and Reporters usually analyze the bugs and points out what causes the bug in natural language in attachments. The Title, Description and Crash file attachments are valuable to us. After analyzing bug information, we have find two types of information that can help developers to locate bug. The first type of information is the attributes written in natural language and the other type is the attributes written in programming language. For the attributes belonging to natural language, feature extraction is done using Bag of Word approach. For example, title and description are generally written in natural language, so word frequency information is considered. The words are assigned probability according to their weighting in classification. For example title," SNMP Query for cempMemBufferMemPoolIndex returns out of range value" ," cempMemBufferMemPoolIndex" should be given more weightage than "SNMP Query".

Attributes of programming language include commands, log events and stack trace decode. For programming language type of attributes, first static analysis of the bug information is done to find out the pattern for their retrieval. For example, event log messages will start with % sign and end with colon (:) like %<feature>:. For bug CSCtn56006, the event log contains messages like

"arf-server59:2011-03-14T15:45:10:%SCRIPT-6-ETEST: %[pname=TRP-Enhanced_MemPool_MIB]: running script Enhanced_Mempool_Mib_en version 1.7".

\*Nov 12 00:30:02.699: %LINEPROTO-5-UPDOWN: Line protocol on Interface GigabitEthernet0/0, changed state to up
\*Nov 12 00:30:02.699: %LINEPROTO-5-UPDOWN: Line protocol on Interface GigabitEthernet0/1, changed state to up
\*Nov 12 00:30:02.699: %LINEPROTO-5-UPDOWN: Line protocol on Interface VoIP-Null0, changed state to up
\*Nov 12 00:30:03.479: %LINK-3-UPDOWN: Interface Serial0/0/0, changed state to down
\*Nov 12 00:30:03.479: %LINEPROTO-5-UPDOWN: Line protocol on Interface IPv6-mpls, changed state to up

Partial Crash File showing Syslog Messages

Pseudo Code for feature Extraction
  Read data from TableI.
    i=0
    For each feature in data
    {
      String=Read string between from and to;
      For each Line in string
      {
        If (separator!="NA" and separator!="pattern- string")
            FEATURE_RESULT[i].add(pattern-string)
        Else
            i++
            FEATURE_RESULT[i].add(pattern-string)
      }
      If (length(FEATURE_RESULT[i])!=0 and no. of pattern>=2)
        Skip next pattern
    }

TABLE I
PATTERN FOR FEATURE EXTRACTION

| S.no | Feature | No. of Pattern | Pattern | From | to | Separator |
|---|---|---|---|---|---|---|
| 1 | Command | 2 | CMD:<feature> | Current Configuration | end | NA |
| 2 | Syslog Messages | 1 | %<feature>: | NA | NA | NA |
| 3 | Tracedecode | 2 | %[0x…]:<feature>+ | NA | NA | Space |
| 4 | Tracdecode | 1 | %[0x…]---><feature>+ | NA | NA | NA |

Commands have two types of pattern (1) start with CMD: like CMD:<feature> (2) Data present between "Current Configuration" and "end" having commands in each line. Consider the following data taken from the crash report of a bug.

CMD: 'no aaa new-model' 19:30:05 EST Sat Nov 11 2006
CMD: 'ip subnet-zero' 19:30:05 EST Sat Nov 11 2006
CMD: 'ip cef' 19:30:05 EST Sat Nov 11 2006
CMD: 'no ip domain lookup' 19:30:05 EST Sat Nov 11 2006
CMD: 'ip domain name tmgcc.csc.com' 19:30:05 EST Sat Nov 11 2006

Partial Crash File showing Command Messages

When the function calls are done the hexadecimal address will be saved in stack not the function name. For analysis the hexadecimal values can be given as input to the decode tool, it will return function names. Here, not only the name of the functions but the order in which they are called is also an important factor. This chunk of function calls retrieved from stack is one of the very important features in classification. Pattern for extraction of these features are listed in Table I.

*C. Feature Selection*

$$IG(f) = -\sum_{k=1}^{m} \Pr(Ck) \log \Pr(Ck) + \Pr(f) \sum_{k=1}^{m} \Pr(Ck|f) \log \Pr(Ck|f) + \Pr(f') \sum_{k=1}^{m} \Pr(Ck|f') \log (\Pr(Ck|f'))$$
(1)

We are considering five feature groups: Title, Description, Syslog Event, Commands and Trace Decode. The features contain some noise also. So to reduce the noise the feature selection is performed using Information gain measure. Information gain is a popular score for feature selection in the field of machine learning. In particular it is used in the C4.5 decision tree inductive algorithm, Yang and Pedersen (1997) have compared different feature selection scores on two datasets and have shown that Information Gain is among the two most effective ones[11]. The information gain of feature f is defined as in (1).

IV. PROBABILISTIC FRAMWORK FOR CLASSIFICATION

In this paper, we are considering Bernoulli and Multinomial Naïve Bayes Model for bug classification purpose, since Naïve Bayes Model is popular for text classification [12]. A Naive Bayes Classifier can be defined as an independent feature model that deals with a simple probabilistic classifier based on Bayes' theorem with strong independence assumptions [13]. There are several models which assume different fitting for Naïve Bayes. The most common models are: Bernoulli Event Model characterized as Boolean weight which uses binary feature occurrences; another one is the multinomial model which uses feature occurrence frequencies. Consider the bug classification into n different classes C = {C1,C2…….,Cn}.The unseen bug(Bi) will be classified using (2) to class with higher posterior probability.

P(Ck,Bi)=P(Bi|Ck)(P(Ck)/P(Bi)) (2)

P(Ck) is the prior probability of class Ck calculated using (3),N is the number of bugs in the training data and Nk is used to denote total number of bugs from training data which belong to class Ck.

P(Ck)=Nk/N (3)

The questions are how do we represent Bi? How do we estimate P(Bi|Ck)? Instead of taking word information as input we are using feature information for bug specific features and for features of natural language type we are considering word information. Words are unigram features but extracted features from bug information may be Bi-gram, Trigram or Multigram. Bug specific features may be a combination of number of words as in Trace Decode and Commands.

*A. Bernoulli Event Model*

In the Bernoulli event model, a bug is represented as a binary vector over the space of features. BVi is the feature vector for the ith bug Bi. We have a vocabulary V containing a set of |V| features each dimension t of the space, where t ε {1,2......|V|}. Dimension t of a bug vector corresponds to feature Ft in the vocabulary. BVit, is either 0 or 1 representing the absence or presence of feature Ft in the ith bug. With such a bug representation, we make the naive Bayes assumption: that the probability of each feature occurring in a bug report is independent of the occurrence of other features in a bug. Then, the probability of a bug given its class from (4) is simply the product of the probability of the attribute values over all feature attributes:

$$P(Bi|Ck) = \prod_{t=1}^{|V|}[BVit\, P(Ft|Ck) + (1 - BVit)(1 - P(Ft|Ck))] \quad (4)$$

The likelihood of each feature is calculated as per (5). Let nk(Ft) be the number of bugs of class Ck in which Ft is observed, and let Nk be the total number of bugs in Ck.

$$P(Ft|Ck) = Nk(Ft)/Nk \quad (5)$$

*B. Multinomial Event Model*

In contrast to the Bernoulli event model, the multinomial model captures feature frequency information in bugs. Consider, for example, Mi is the multinomial model feature vector for the ith bug data Bi. Mit, is the number of times feature Ft occurs in bug data Bi; ni= ∑t Mit the total number of features in Bi. In the multinomial model, a bug is an ordered sequence of feature events, drawn from the same vocabulary V. We assume that the lengths of bugs are independent of class. We again make a similar naive Bayes assumption: that the probability of each feature event in a bug is independent of the feature's context and position in the document. P(Ft|Ck) is estimated using word frequency information from the multinomial model feature vectors.
Generation of bugs is modeled by repeatedly drawing features from a multinomial distribution.

$$P(Mi|Ck) = \frac{n!}{\prod_{t=1}^{|V|} Mit!} \prod_{t=1}^{|V|} P(Ft|Ck)^{Mit} \quad (6)$$

If comparing likelihoods of the same bug for different classes (e.g. P(Mi|Ck) vs P(Mi|Cj)), then

$$P(Mi|Ck) \propto \prod_{t=1}^{|V|} P(Ft|Ck)^{Mit} \quad (7)$$

If N is the total number of bugs then Estimate P(Ft|Ck) using (8) as relative frequency of Ft in bugs of class Ck with respect to the total number of features in bug data of that class. zik shows the presence or absence of feature Ft in the bug feature vector.

$$P(Ft|Ck) = \frac{\sum_{i=1}^{N} Mit\, zik}{\sum_{s=1}^{|V|} \sum_{i=1}^{N} Mis\, zik} \quad (8)$$

*A. Dataset Information*

The data from six different categories os, bgp, ip, ipv6, aaa, snmp are collected from the site of one of the networking based organization. The training data contains 1000-1500 bugs from each category. Five different types of features are extracted by static analysis and pattern matching. The Syslog Event contain around 30,000 Syslog messages, for Commands The vocabulary size is around 600 commands, for Title and Description word frequency data is taken and their respective vocabulary sizes are around 9000 and 30000. For Trace decode, there are 400 chunks available for the classification purpose.

*B. Performance Measure*

In classification system, the terms true positives, true negatives, false positives, and false negatives are used to compare the results of the classifier under test with known external output. The terms positive and negative refer to the classifier's prediction and the terms true and false refer to whether that prediction corresponds to the external output. To evaluate the performance of bug classification system, we are using four standard measures Precision, Recall, Accuracy and F-Measure.

For Precision and Recall we used the standard definitions,

$$\text{Precision} = \frac{\text{Categories found and correct}}{\text{TotalCategories Correct}} \quad (9)$$

$$\text{Recall} = \frac{\text{Categories found and correct}}{\text{TotalCategories Found}} \quad (10)$$

The Precision and Recall will be calculated for all the categories(class) and Accuracy is taken to assign the Precision and Recall values for the classification. The measures calculated for classification algorithms Bernoulli and Multinomial is shown in Table II. The measures are taken across all the feature groups under consideration as shown in Table III and IV. The two other measures Accuracy and F-Measure is also calculated. F-Measure is defined as the harmonic of Precision and Recall.

$$\text{F-Measure} = \frac{2.\,\text{Precision} * \text{Recall}}{\text{Precision} + \text{Recall}} \quad (11)$$

$$\text{Accuracy} = \frac{\text{Bug Categorized Correctly}}{\text{Total Number of Bugs}} \quad (12)$$

*C. Results and Discussion*

Bug classification is entirely different from text classification. The classes which are considered in bug classification (all bugs of network type) are on the basis of network protocols like IPV4 (referred as IP in literature), IPv6, SNMP, BGP and OS. Since these classes have many common characteristic and do not have fixed boundaries like what we have for text classification. The feature selection has been done on the basis of Information Gain Measure.
Experimental results for two class classification using word information taken from the bug data gives an accuracy of around 60% and 78% for Bernoulli and Multinomial respectively. But as we move to Multiclass classification, it gives an accuracy of less than 15%. According to us, using bug specific features like Title, Description, Syslogs, Commands and Trace Decode accuracy of classification can be increased. The experimental results show the effect of applying Bernoulli and Multinomial Model to the bug data using bug specific features. The Precision, Recall, Accuracy and F-Measure values for both models are as follows:

TABLE II
AVERAGE CLASSIFIER PERFORMANCE FOR BUG CLASSIFICATION

| Classification Model | Precision | Recall | F-Measure | Accuracy |
|---|---|---|---|---|
| Bernoulli | 0.4026 | 0.8163 | 0.5094 | 0.4171 |
| Multinomial | 0.40634 | 0.81342 | 0.51226 | 0.4194 |

In our experiments we have implemented Multinomial and Bernoulli Models on all feature groups.

TABLE III
PARAMETER VALUES FOR BERNOULLI MODEL

| Feature Group | Precision | Recall | F-Measure | Accuracy |
|---|---|---|---|---|
| Title | 0.5202 | 0.9867 | 0.6644 | 0.5205 |
| Description | 0.5432 | 0.9806 | 0.6767 | 0.5437 |
| Syslogs | 0.3737 | 0.8675 | 0.4833 | 0.3798 |
| Commands | 0.3372 | 0.6910 | 0.4348 | 0.3695 |
| Traces | 0.2385 | 0.5555 | 0.2878 | 0.2719 |

The comparison of F-measure values for all the feature groups between Bernoulli and Multinomial Models is shown in Table III and IV. We can observe that Bernoulli Model is giving good results when compared to the Multinomial model. And for Trace Decode the results are being similar. After assignment of priorities to the feature groups, the feature groups are arranged. The new unseen bug will go through the feature group checking in the order of the priorities assigned to them. At the time of classification considering Description and Syslogs feature groups, Multinomial model is applied. For rest of the feature groups Bernoulli model is referred. The overall accuracy of the classification is found to be 55% after applying the mentioned sequence.

TABLE IV
PARAMETER VALUES FOR MULTINOMIAL MODEL

| Feature Group | Precision | Recall | F-Measure | Accuracy |
|---|---|---|---|---|
| Title | 0.4437 | 0.9085 | 0.5736 | 0.4443 |
| Description | 0.5928 | 0.9840 | 0.7244 | 0.5924 |
| Syslogs | 0.4354 | 0.8821 | 0.5482 | 0.4403 |
| Commands | 0.3213 | 0.7370 | 0.4273 | 0.3484 |
| Traces | 0.2385 | 0.5555 | 0.2878 | 0.2719 |

## VI. CONCLUSION AND FUTURE WORK

The experimental results show that applying Bayesian model for classification of bugs using word information as feature is reliable for two classes. But it does not give any proper reason for the classification and cannot be used for multiclass classification. It can be concluded from the results that we need to go into semantics of bug information. We had successfully extracted some of the bug specific features. According to the results, Bernoulli and Multinomial Models using bug specific features give better accuracy compared to word information. In future we will extract some more features and apply some more apt classification algorithm such as SVM and neural network. Using the new technique and specific features we will try to improve the accuracy of classification .

## VII. .REFERENCES

[1] G. O. Young, "Synthetic structure of industrial plastics (Book style with paper title and editor)," in *Plastics*, 2nd ed. vol. 3, J. Peters, Ed. New York: McGraw-Hill, 1964, pp. 15–64.
[2] W.-K. Chen, *Linear Networks and Systems* (Book style). Belmont, CA: Wadsworth, 1993, pp. 123–135.
[3] H. Poor, *An Introduction to Signal Detection and Estimation*. New York: Springer-Verlag, 1985, ch. 4.
[4] B. Smith, "An approach to graphs of linear forms (Unpublished work style)," unpublished.
[5] E. H. Miller, "A note on reflector arrays (Periodical style—Accepted for publication)," *IEEE Trans. Antennas Propagat.*, to be published.
[6] J. Wang, "Fundamentals of erbium-doped fiber amplifiers arrays (Periodical style—Submitted for publication)," *IEEE J. Quantum Electron.*, submitted for publication.
[7] C. J. Kaufman, Rocky Mountain Research Lab., Boulder, CO, private communication, May 1995.
[8] Y. Yorozu, M. Hirano, K. Oka, and Y. Tagawa, "Electron spectroscopy studies on magneto-optical media and plastic substrate interfaces(Translation Journals style)," *IEEE Transl. J. Magn.Jpn.*, vol. 2, Aug. 1987, pp. 740–741 [*Dig. 9th Annu. Conf. Magnetics* Japan, 1982, p. 301].
[9] M. Young, *The Techincal Writers Handbook*. Mill Valley, CA: University Science, 1989.
[10] J. U. Duncombe, "Infrared navigation—Part I: An assessment of feasibility (Periodical style)," *IEEE Trans. Electron Devices*, vol. ED-11, pp. 34–39, Jan. 1959.
[11] S. Chen, B. Mulgrew, and P. M. Grant, "A clustering technique for digital communications channel equalization using radial basis function networks," *IEEE Trans. Neural Networks*, vol. 4, pp. 570–578, July 1993.
[12] R. W. Lucky, "Automatic equalization for digital communication," *Bell Syst. Tech. J.*, vol. 44, no. 4, pp. 547–588, Apr. 1965.
[13] S. P. Bingulac, "On the compatibility of adaptive controllers (Published Conference Proceedings style)," in *Proc. 4th Annu. Allerton Conf. Circuits and Systems Theory*, New York, 1994, pp. 8–16.
[14] G. R. Faulhaber, "Design of service systems with priority reservation," in *Conf. Rec. 1995 IEEE Int. Conf. Communications,* pp. 3–8.
[15] W. D. Doyle, "Magnetization reversal in films with biaxial anisotropy," in *1987 Proc. INTERMAG Conf.*, pp. 2.2-1–2.2-6.

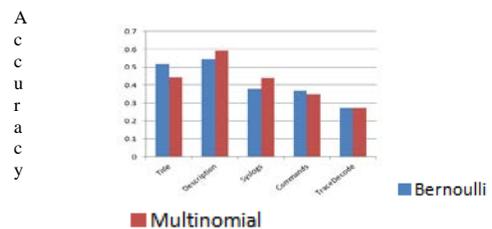

Fig. 1  Accuracy Comparison of Feature Groups